**Corrigendum: Shear and normal stress measurements in non-Brownian monodisperse and bidisperse suspensions in J. Rheol. 60(2), 289-296.**

Chaiwut Gamonpilas,[1, 2, a)] Jeffrey F. Morris,[2, 3, b)] and Morton M. Denn[2, 3, c)]

[1)] MTEC, National Science and Technology Development Agency (NSTDA), 114 Thailand Science Park, Pahonyothin Road, Khlong 1, Khlong Luang, Pathum Thani 12120, Thailand.
[2)] Benjamin Levich Institute, City College of New York, CUNY, New York, NY 10031, USA.
[3)] Department of Chemical Engineering, City College of New York, CUNY, New York, NY 10031, USA.

a) Author to whom correspondence should be addressed; electronic mail: chaiwutg@mtec.or.th
b) Electronic mail: morris@ccny.cuny.edu
c) Electronic mail: denn@ccny.cuny.edu

There was an error in data reduction, resulting in incorrect values for the normal stress differences $N_1$ and $N_2$ shown in Figs. 7-10, and the corrected figures are shown here. In particular, the algebraic sign of $N_1$ is changed, as are the relative magnitudes of $N_1$ and $N_2$. The negative values of $N_1$ for these non-shear-thickening suspensions are larger in magnitude than those reported by other workers, but both $N_1$ and $N_2$ are in general agreement with the accelerated Stokesian Dynamics calculations of Sierou and Brady [1].

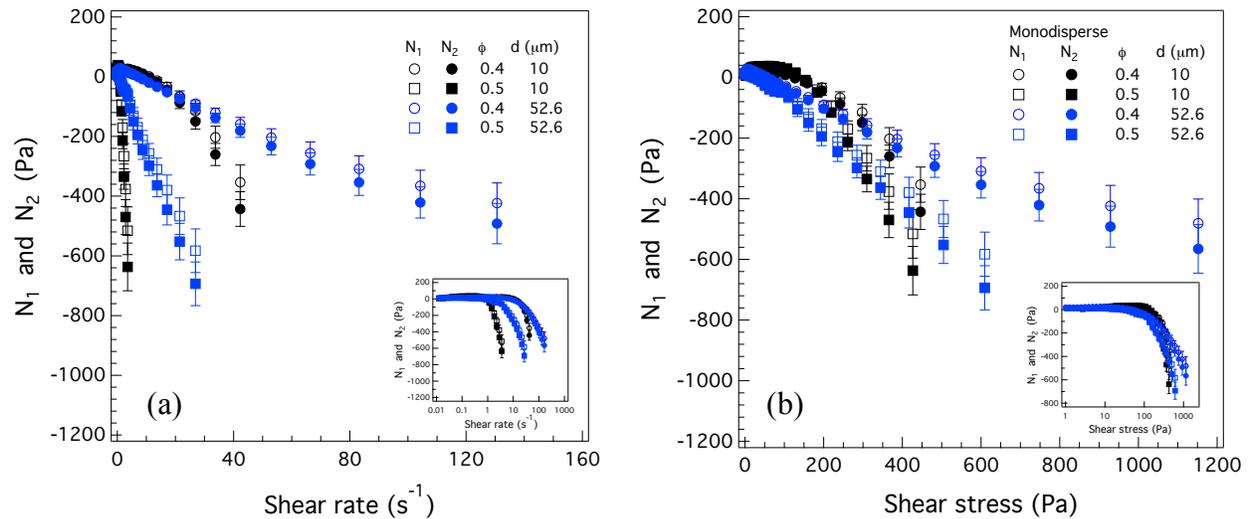

FIG. 7. Normal stress differences $N_1$ and $N_2$ of monodisperse suspensions at volume fractions of 0.4 and 0.5 for both 10 and 52.6 μm particles as functions of (a) shear rate and (b) shear stress (Insets are the same data on a semilog scale).

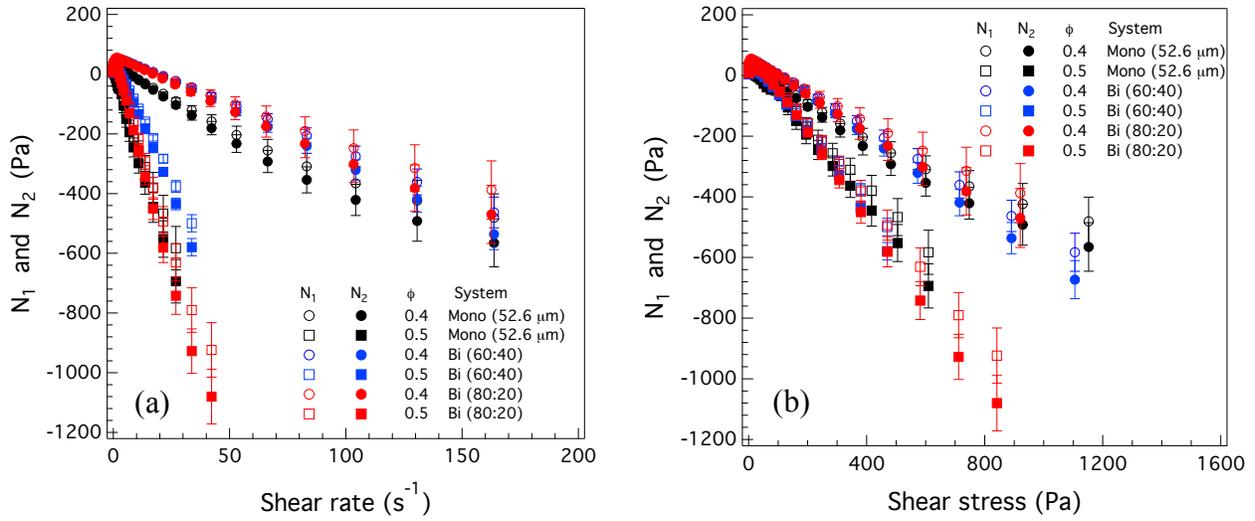

FIG. 8. Normal stress differences $N_1$ and $N_2$ of mono- and bidisperse suspensions at volume fractions of 0.4 and 0.5 as functions of (a) shear rate and (b) shear stress.

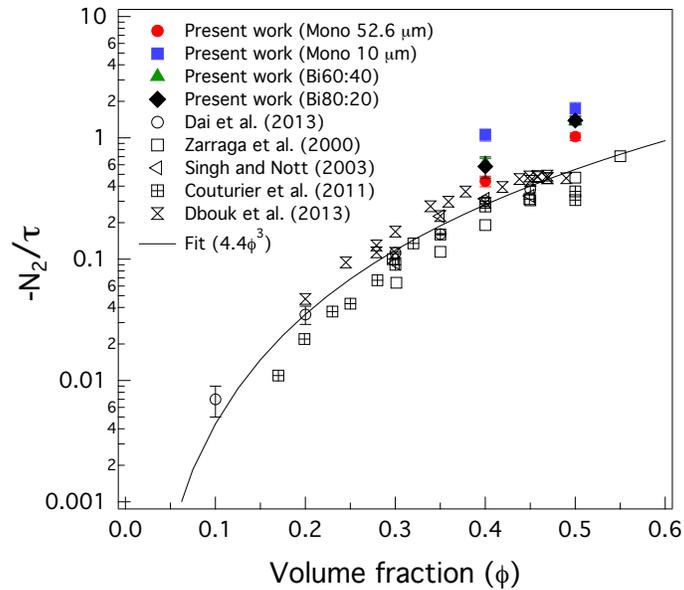

FIG. 9. Comparison of present results for $-N_2/\tau$ with experimental results from other studies.

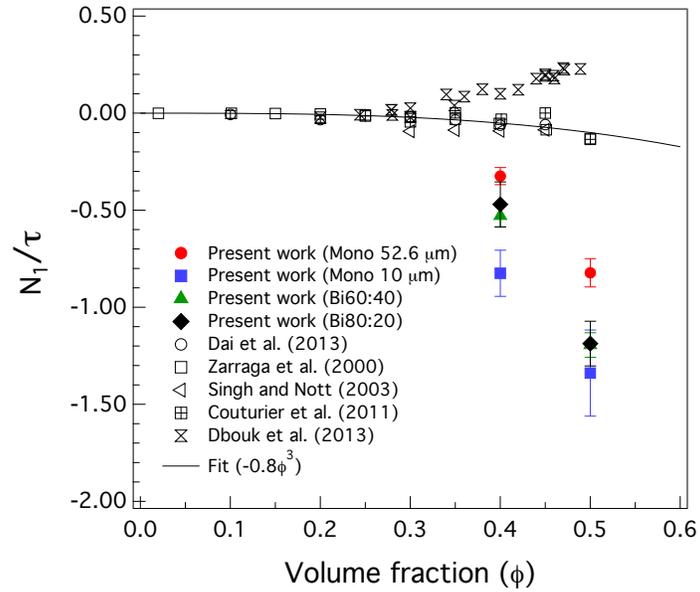

FIG. 10. Comparison of present results for $N_1/\tau$ with experimental results from other studies.

# Shear and Normal Stress Measurements in Non-Brownian Monodisperse and Bidisperse Suspensions

Chaiwut Gamonpilas,[1, 2, a)] Jeffrey F. Morris,[2, 3, b)] and Morton M. Denn[2, 3, c)]

[1)] MTEC, National Science and Technology Development Agency (NSTDA), 114 Thailand Science Park, Pahonyothin Road, Khlong 1, Khlong Luang, Pathum Thani 12120, Thailand.

[2)] Benjamin Levich Institute, City College of New York, New York, NY 10031, USA.

[3)] Department of Chemical Engineering, City College of New York, New York, NY 10031, USA.

a) Author to whom correspondence should be addressed; electronic mail: chaiwutg@mtec.or.th

b) Electronic mail: morris@ccny.cuny.edu

c) Electronic mail: denn@ccny.cuny.edu

(Dated: 1 December 2015)

## SYNOPSIS

We have measured the viscometric functions of mono- and bimodal non-colloidal suspensions of PMMA spheres in a density-matched aqueous Newtonian suspending fluid using parallel-plate and cone-and-plate rheometry for particle volume fractions in the range 0.20 to 0.50. Cone-and-plate normal stress measurements employed the method of Marsh and Pearson, in which there is a finite gap between the cone tip and the plate. The monodisperse suspensions showed an unexpected particle size dependence of the viscometric functions, with the viscosity increasing with decreasing particle size. Normal stresses were very small in magnitude and difficult to measure at volume fractions below 0.30. At higher concentrations, $N_2$ was negative and much larger in magnitude than $N_1$, for which the algebraic sign was positive over most of the shear rate range for the monodisperse suspensions but indeterminate and possibly negative for the bimodal suspensions. The normal stresses were insensitive to bidispersity when plotted as functions of the shear stress at each volume fraction.

*Keywords:* Suspensions, non-colloidal, bidisperse, normal stresses, size dependence

# I. INTRODUCTION

Suspensions of solid particles are ubiquitous in a wide range of industrial applications, including foods, pharmaceuticals, cosmetics, plastics, fertilizers, construction, and oil well engineering. Natural hazards such as landslides and mud and lava flows also contain particle suspensions. The rheological properties of these suspensions are complex and highly dependent on a number of factors, including fluid properties and particle size, shape, and volume fraction. Knowledge of the suspension rheology is important for understanding and manipulating flow behavior in order to ensure optimal performance [Wagner and Brady, 2009; Mewis and Wagner, 2011; Denn and Morris, 2014]. In this work, we study the stresses generated in viscometric flow of monodisperse and bidisperse suspensions of non-Brownian spheres in a Newtonian suspending fluid, namely the shear stress $\sigma_{12}$ and the normal stress differences $N_1 = \sigma_{11} - \sigma_{22}$ and $N_2 = \sigma_{22} - \sigma_{33}$. Here, the standard convention is used for the velocity (1), velocity gradient (2) and vorticity (3) directions.

The effective viscosity of monodisperse suspensions has been described by many semi-empirical models, including those of Mooney [Mooney, 1951], Krieger and Dougherty [Krieger and Dougherty, 1959], Maron and Pierce [Maron and Pierce, 1956] Quemada [Quemada, 1977], Zarraga and co-workers [Zarraga, Hill, and Leighton, 2000], and the recent model of Mendoza and Santamaria-Holek [Mendoza and Santamaria-Holek, 2009]. These models all include the maximum packing fraction $\phi_m$ as a parameter; $\phi_m$ is the volume fraction at which the system is presumed to jam and fluidity ceases. Values of the maximum packing fraction have been reported over a range of values of 0.49-0.64, although in practice it is not straightforward to make suspensions flow above 0.55 [Guazzelli and Morris, 2011]. For bidisperse and polydisperse suspensions, the effective viscosity is reduced significantly and the maximum packing increases relative to a monodisperse suspension [Farris, 1968; Chong, Christainsen and Baer, 1971; Poslinski *et al.*, 1988; Shapiro and Probstein, 1992; Chang and Powell, 1994; He and Ekere, 2001]. When only hydrodynamic forces act on the particles, it is a result of dimensional analysis that the shear stress should exhibit a linear dependence on the magnitude of the shear rate in a Stokesian suspension. As a consequence, the shear viscosity is expected to be Newtonian. The deviation from Newtonian behavior of non-colloidal Stokesian suspensions was first discussed by Bagnold [1954], who demonstrated a radial normal stress that varied linearly with the shear rate in a cylindrical-Couette geometry. Gadala-Maria and Acrivos [1980] subsequently measured $N_1 - N_2$ using a rheometer equipped with a parallel plate geometry and showed that the value was directly proportional to the shear stress, but experimental reproducibility was difficult to achieve. Singh and Nott [2003] measured the normal stresses of non-colloidal suspensions using cylindrical-Couette and parallel-plate geometries. They found that both $N_1$ and $N_2$ are negative and vary linearly with shear rate. More recent measurements by a number of research groups have established that $N_2$, which is negative, is the dominant normal stress difference in non-Brownian suspensions of spheres with volume fractions greater than about 0.3, in contrast to flexible polymers, where the dominant normal stress difference is $N_1$, which is positive. However, the sign and magnitude of $N_1$ is still inconclusive; the recent study by Dbouk, Lobry, and Lemaire [2013] reports a positive $N_1$, whereas other studies report negative $N_1$ values [Zarraga *et al.*, 2000; Dai *et al.*, 2013]. A comprehensive review of non-Brownian suspension rheology can be found in Denn and Morris [2014].

Most studies have been focused on monodisperse systems, and little is known about the rheological properties of bi- or polydisperse suspensions of spheres in a Newtonian suspending fluid, particularly the normal stresses. Examination of the normal stresses in bidisperse suspension is a focus of the present work.

## II EXPERIMENTAL

**A. Materials and preparation of suspensions**

The particles were 10.10 ± 1.12, 19.6 ± 0.27 and 52.60 ± 9.80 μm poly (methyl methacrylate) (PMMA) spheres with a density of approximately 1,200 kg/m$^3$, purchased from Microbeads AS (Norway). A suspending fluid matching the refractive index and density of the PMMA particles was prepared following Krishnan *et al.* [1996]: 77.93% Triton X-100, 9.01% anhydrous zinc chloride, and 13.06% water (by weight). The resulting fluid was Newtonian, with a viscosity of 1.01 ± 0.01 Pas at 25 °C (Fig. 1); the temperature dependence from 20 to 40 °C is fit by $\eta_0$ =1.01e$^{-0.07(T-25)}$ (Fig. 1 inset). Mono- and bidisperse suspensions were prepared at volume fractions from $\phi$ = 0.2 to 0.5. For the bidisperse system, volume ratios of large ($\phi_l$) to small ($\phi_s$) particles of 60:40 and 80:20 were studied. The dispersions were prepared by manually stirring to achieve homogeneity, after which they were left overnight to allow any air bubbles to rise; air bubbles were skimmed from the top surface before loading a sample into the rheometer.

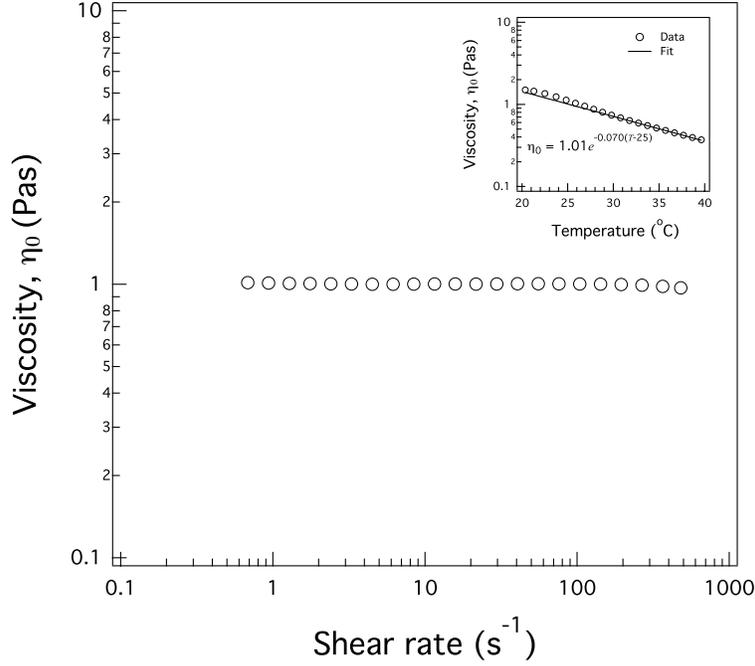

FIG. 1. Viscosity of suspending fluid as a function of shear rate and temperature (inset).

**B. Rheology**

An ARES-G2 strain controlled rheometer with parallel-plate and cone-and-plate geometries was used to measure the viscometric properties of the suspensions. This rheometer is equipped with a force rebalance transducer (FRT) that provides a normal force range of 0.001-20 N. Temperature control is achieved through the Advanced Peltier System (APS), with temperature measured by a thermocouple in contact with the plate.

The viscosity $\eta$ and normal stress difference $N_1 - N_2$ were measured in a parallel-plate configuration; the parallel-plate diameter was 50 mm, and the gap used in all experiments was 1.25 mm, which was large enough to eliminate wall slip effects [Zarraga, Hill, and Leighton, 2000]. The viscometric functions were calculated using the following equations:

$$\eta(\dot{\gamma}_R) = \frac{Mh}{2\pi\Omega R^4}\left(3 + \frac{d\ln M}{d\ln \dot{\gamma}_R}\right), \tag{1}$$

$$N_1(\dot{\gamma}_R) - N_2(\dot{\gamma}_R) = \frac{F}{\pi R^2}\left(2 + \frac{d\ln F}{d\ln\dot{\gamma}_R}\right). \tag{2}$$

Here, $M$ is the torque, $F$ is the total thrust, $R$ is the radius, $h$ is the gap spacing, $\Omega$ is the rotational speed, and $\dot{\gamma}_R = R\Omega/h$.

Classical cone-and-plate rheometry cannot be used because of the large particle sizes, and we followed the procedure of Marsh and Pearson [1968] that employs a finite gap between the cone tip and the plate to obtain a second relation between the two normal stress differences, which is given at a gap spacing $C$ by

$$N_2(\dot{\gamma}_R) = \left[\frac{F}{\pi R^2}\left(-\frac{d\ln F}{d\ln C} + 2\right) + N_1(\dot{\gamma}_R)\right]\left[1 + \frac{R}{C}\tan\theta\right], \tag{3}$$

where $\theta$ is the cone angle. The cone was 25 mm in diameter, with 0.0201 rad angle and a 40 μm truncated gap. Gap spacings of 0.6 and 0.7 mm were used, and $\partial\ln F/\partial\ln C$ was evaluated as the difference $\Delta\ln F/\Delta\ln C$.

$N_1$ and $N_2$ can thus be obtained by measuring the normal thrust as a function of shear rate in both geometries. The effects of inertia and secondary flows were negligible, as the largest particle Reynolds number was less than $10^{-5}$. The impact of Brownian motion was also negligible, as the Péclet number was very high.

In all experiments, samples were loaded using a gap control speed of 0.1 mm/s. After loading, the excess was trimmed away from the geometry edge. A preshear of 1 s$^{-1}$ was then applied for 60 s to remove any loading history and to set the initial state of the sample. At the end of the preshear, the sample was allowed to rest for another 60 s prior to performing a viscometric measurement. The measurement of the viscosity and normal stress differences was performed using ramp-up flow sweeps from 0.01 to 1000 s$^{-1}$. Data points were acquired by allowing the sample to equilibrate for 60 s, after which the viscosity and normal force were recorded over the following 900 s.

## III RESULTS AND DISCUSSION

### A. Viscosity

The viscosity of monodisperse PMMA suspensions from $\phi = 0.2$ to $0.5$ for both 10 and 52.6 μm particles is shown in Fig. 2(a). Shear thinning is observed for particle volume fractions of 0.3 and greater [cf. Zarraga, Hill, and Leighton, 2000; Stickel and Powell, 2005; Tanner, 2015]. Reproducibility is quite good, as shown in Fig. 2(b) for the 10 μm particles; similar reproducibility was observed for the 52.6 μm particles (not shown). The slight shear thickening observed at $\phi = 0.5$ for the 52.6 μm particles is reproducible and has been observed in previous studies [Zarraga, Hill, and Leighton, 2000; Dai et al., 2013; Tanner, 2015]. A striking feature is that the viscosity at volume fractions greater than 0.3 is dependent on the particle size, with the smaller particles exhibiting a higher viscosity. The particle size dependence is clearly seen in Fig. 3, where the relative viscosity is plotted as a function of $1/d$ for $\phi = 0.2$, $0.3$, and $0.4$.

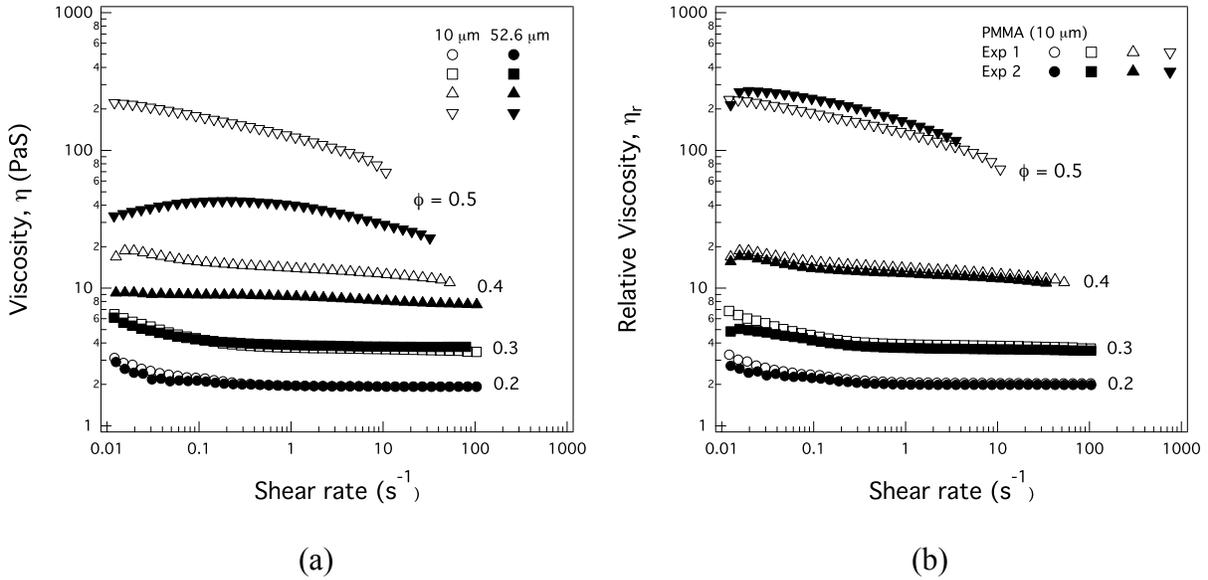

FIG. 2. (a) Viscosity of monodisperse suspensions prepared from 10 and 52.6 μm particles and (b) experiment repeatability of 10 μm monodisperse suspensions for $\phi = 0.2$ to $0.5$.

Dependence of the suspension viscosity on particle size is unexpected; it is not seen, for example, in the classic study by Lewis and Nielsen [1968], in which glass particles with diameters ranging from 5 to 105 μm were used. Consequently, we carried out viscosity measurements on suspensions of glass spheres with diameters of 41.5 ± 3.6 and 122.6 ± 8.7 μm (see supplemental data in Appendix). No significant size dependence was evident, consistent with previous studies of similar suspensions [Lewis and Nielsen, 1968; Shapiro and Probstein, 1992; Zarraga, Hill, and Leighton, 2000], suggesting that the size dependence observed here will vary from system to system. The simulations of Seto *et al.* (2013) and Mari *et al.* (2014) show the importance of particle-particle repulsive forces in concentrated suspensions of non-Brownian spheres, and such forces for different particle-fluid systems may be a factor in the observed size dependence. The size dependence is an interesting and important open question that is closely related conceptually to the size dependence of the rheology of filled polymers [Mermet-Guyennet *et al.*, 2015]. While mechanisms are undoubtedly different, in both cases there is a need to understand the interactions that lead to a loss of scale invariance.

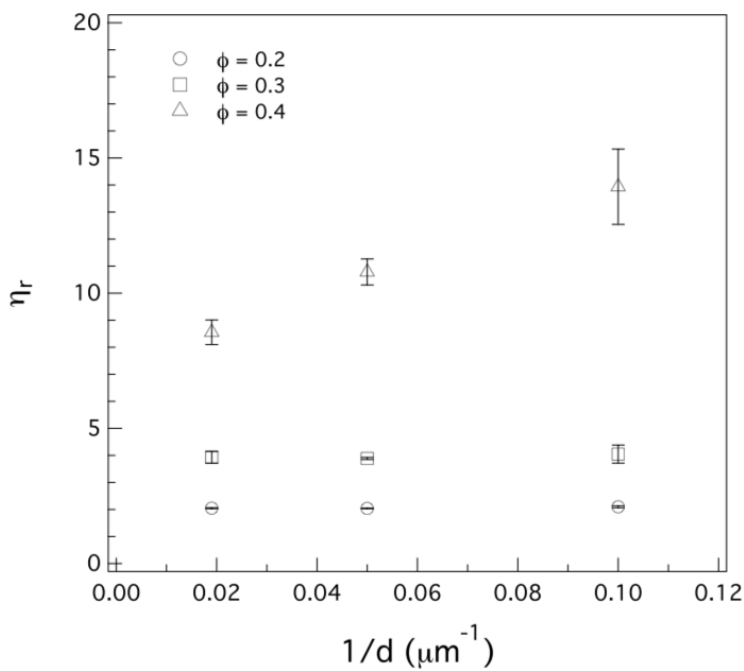

FIG. 3. Size dependence of the relative viscosity for 10, 20 and 52.6 μm PMMA particles at $\phi$ = 0.2, 0.3 and 0.4. Viscosity data are averages between shear rates of 0.1 and 10 s$^{-1}$.

The relative viscosities of monodisperse suspensions of 10 and 52.6 μm PMMA particles are shown in Fig. 4, together with bidisperse suspensions of these particle sizes at volume ratios $\phi_l:\phi_s$ of 60:40 and 80:20. The solid line is the correlation of Zarraga, Hill, and Leighton [2000], $\eta_r = e^{-2.34\phi} (1 - \phi/\phi_m)^{-3}$, with the maximum packing parameter $\phi_m$ set to 0.627. The correlation is a good fit to the glass data of Lewis and Nielsen [1968] for all particle sizes and to the data for the 52.6 μm PMMA suspensions, whereas the data for the 10 μm PMMA suspensions are significantly higher for volume fractions of $\phi = 0.4$ and above. The bidisperse data lie below the correlation line, indicating that the jamming point is increased with particle bidispersity. This behavior for the bidisperse system is in good agreement with most previous experiments [Farris, 1968; Shapiro and Probstein, 1992; Gondret and Petit, 1997; He and Ekere, 2001; Shewan and Stokes, 2015]. Simulations of bidisperse suspensions by Chang and Powell (1993) suggest that the reduction of viscosity is due to a disruption of clustering that is seen in monodisperse suspensions. Sengun and Probstein [1989] developed a model for the fluidity limit of bidisperse suspensions that gives a somewhat lower value in the monodisperse limit than is obtained from the correlation of Zarraga and coworkers. Recently, Qi and Tanner [2011, 2012] proposed a model to describe the random close packing and relative viscosity of bimodal and multi-model suspensions. Their model adopts the formula of Mendoza and Santamaria-Holek [2009] for the relative viscosity, $\eta_r = [1 - \phi/(1 - c\phi)]^{-5/2}$, with $c = (1 - \phi_m)/\phi_m$. The fits of this model to the monodisperse PMMA suspensions are shown in Fig. 4, showing a relatively good fit with lower maximum packing compared to the Zarraga model and decreasing maximum packing fraction with decreasing particle size. The fluidity limits for the bidisperse PMMA suspensions calculated from the viscosity measurements using the Mendoza and Santamaria-Holek model are shown in Fig. 5, together with Shapiro and Probstein's [1992] fluidity limit data for bidisperse glass suspensions with particle size ratios of 2:1 and 4:1. In addition, the random close packing of bimodal suspensions derived from Qi and Tanner's model [Qi and Tanner, 2011] are shown on the same plot. It is evident that the maximum packing fraction curves derived from the viscosity data are qualitatively similar, with the largest value found roughly at a fraction 0.6 of the large particles. The agreement of maximum packing fraction between both sets of experimental data and the model is good, with the errors under 4%. However, one difference is that the monodisperse limits at the two ends of our PMMA data are not the same because of the size dependence noted above.

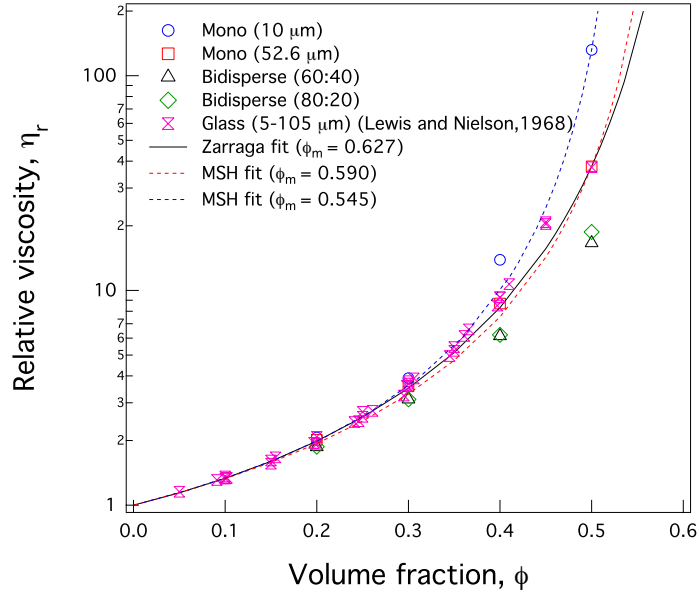

FIG. 4. Relative viscosity of monodisperse suspensions for 10 and 52.6 μm PMMA particles and bidisperse suspensions of these particle sizes at volume ratios $\phi_l:\phi_s$ of 60:40 and 80:20. All data were taken at a shear rate of 1 s$^{-1}$.

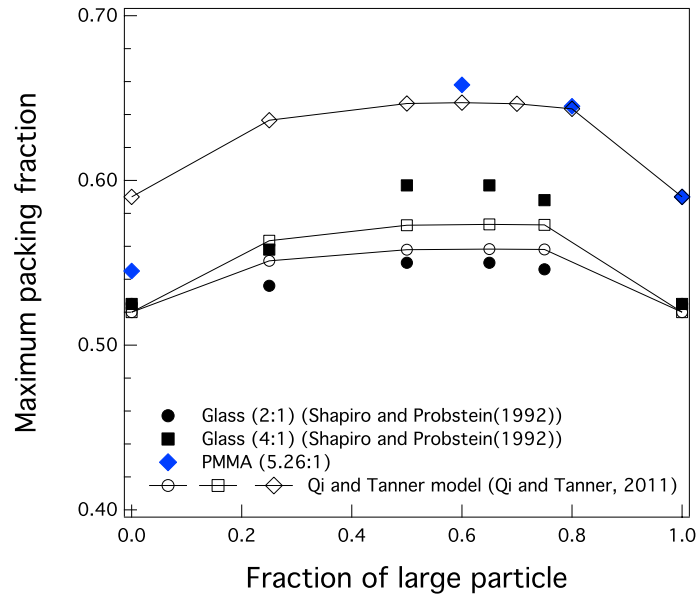

FIG. 5. Maximum packing fractions for PMMA bidisperse suspensions with a particle size ratio of 5.26:1 using the Mendoza and Santamaria-Holek model to fit the relative viscosity data, shown together with the lower bound fluidity limit fractions of glass particles with particle size ratios of 2:1 and 4:1 taken from Shapiro and Probstein [1992] and those calculated using Qi and Tanner's model [Qi and Tanner, 2011].

## B. Normal stresses

The normal stress differences were obtained from total force measurements in parallel-plate and cone-and-plate geometries using Eqs. (2) and (3), with gap sizes of 0.6 and 0.7 mm in the cone-and-plate geometry to avoid jamming at the cone apex. As shown in Fig. 6, the measured normal thrust in the cone-and-plate configuration was very small and almost independent of the gap spacing at shear rates below 10 s$^{-1}$, whereas a slightly higher normal force was obtained for the larger gap at higher shear rates. As a result, the term $\partial \ln F / \partial \ln C$ in Eq. 1 was taken to be zero below 10 s$^{-1}$ but had a value between 0 and 1 for higher shear rates. Because only two points were used, the value of the derivative was the same for calculations with both gaps. The normal stress differences $N_1$ and $N_2$ were obtained by solving Eqs. (2) and (3) simultaneously, with the results differing slightly depending on which gap was used for the cone-and-plate data.

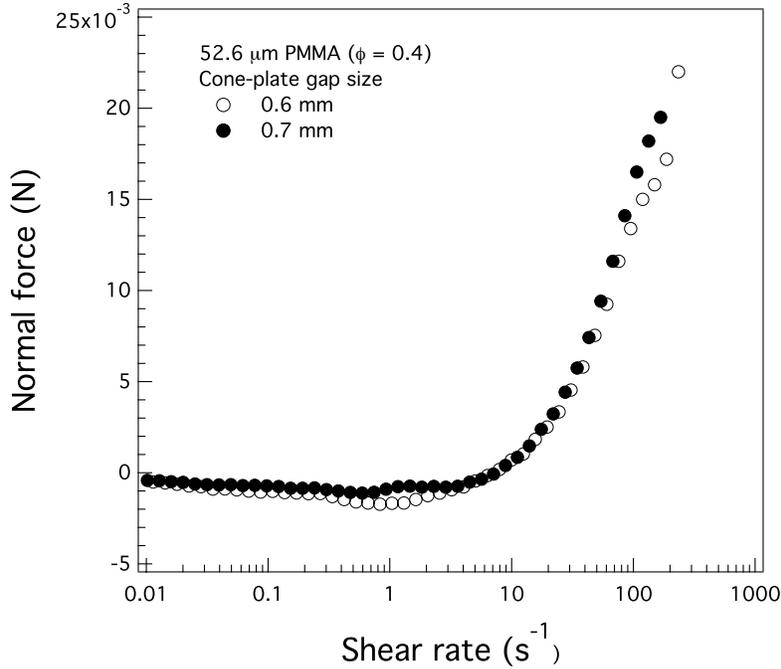

FIG. 6. Measured normal thrust as a function of shear rate for a monodisperse suspension (52.6 μm PMMA particles at $\phi = 0.4$) at two gap sizes of 0.6 and 0.7 mm.

Figures 7(a) and 7(b) show the normal stress differences for monodisperse suspensions at $\phi = 0.4$ and 0.5 for both 10 and 52.6 μm PMMA particles as functions of shear rate and shear stress, respectively. The data points represent the average of the data from the two gaps, while the error bars show the difference between the two. The normal stresses for smaller volume fractions are not resolvable, consistent with prior studies [Zarraga, Hill, and Leighton, 2000; Singh and Nott, 2003; Boyer, Pouliquen, and Guazzelli, 2011], and they are not resolvable at shear stresses below 10 Pa for these volume fractions because of system sensitivity. The second normal stress difference, $N_2$, is negative and larger in magnitude than $N_1$, for which the sign is inconclusive for some of the data. There is a significant size effect when the normal stress data are plotted versus shear rate, but less so when plotted versus shear stress.

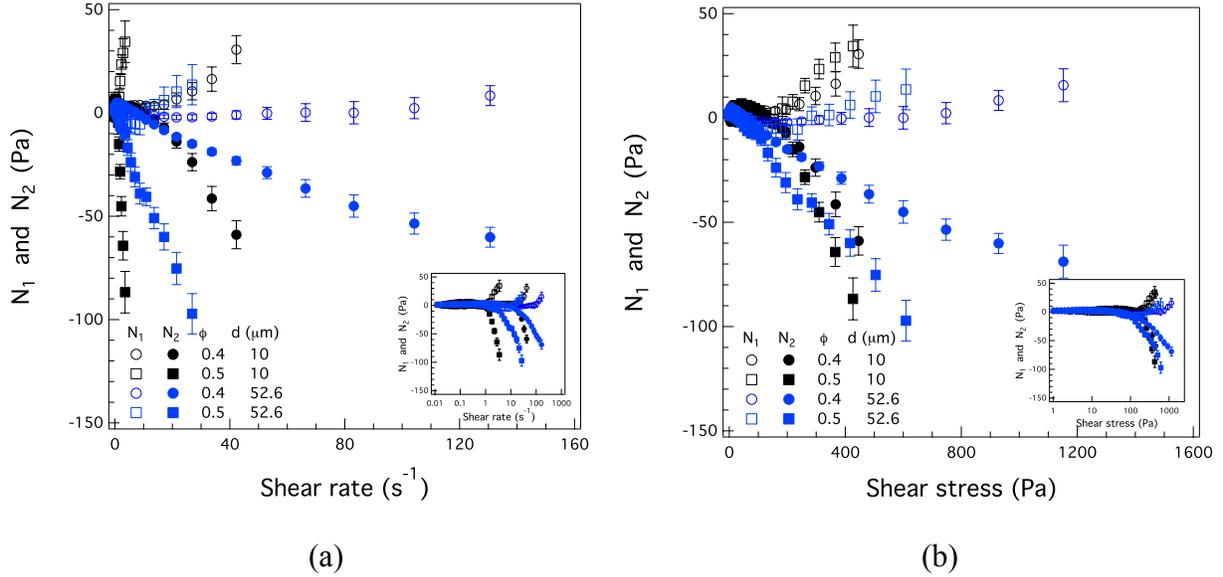

FIG. 7. Normal stress differences $N_1$ and $N_2$ of monodisperse suspensions at volume fractions of 0.4 and 0.5 for both 10 and 52.6 µm particles as functions of (a) shear rate and (b) shear stress (Insets are the same data on a semi-log scale).

Negative $N_2$ is predicted by theory and computation [e.g., Brady and Morris, 1997; Sierou and Brady, 2002; Mari *et al.*, 2014] and is observed experimentally [Singh and Nott, 2003; Dai *et al.*, 2013; Dbouk, Lobry, and Lemaire, 2013; Tanner, 2015]. There is no general agreement about the algebraic sign or magnitude of $N_1$. Calculations using accelerated Stokesian dynamics predict negative $N_1$ with a magnitude comparable to $N_2$ [Sierou and Brady, 2002]. Particle-level simulations incorporating particle-particle friction [Seto *et al.*, 2013; Mari *et al.*, 2014] generally show a small $N_1$, with a possible transition from negative to positive at the shear thickening transition. Many experiments report negative $N_1$ [Zarraga *et al.*, 2000; Singh and Nott, 2003; Dai *et al.*, 2013], but Dbouk, Lobry, and Lemaire [2013] found small positive $N_1$, which is in agreement with the present study.

Figure 8 shows the normal stress differences $N_1$ and $N_2$ of the two bidisperse systems (volume ratios $\phi_l:\phi_s$ of 60:40 and 80:20), together with those for the monodisperse system with 52.6 μm particles. $N_2$ is negative and larger in magnitude than $N_1$, as with the monodisperse system, and the sign of $N_1$ is not clear. The normal stresses as functions of shear stress are insensitive to bidispersity.

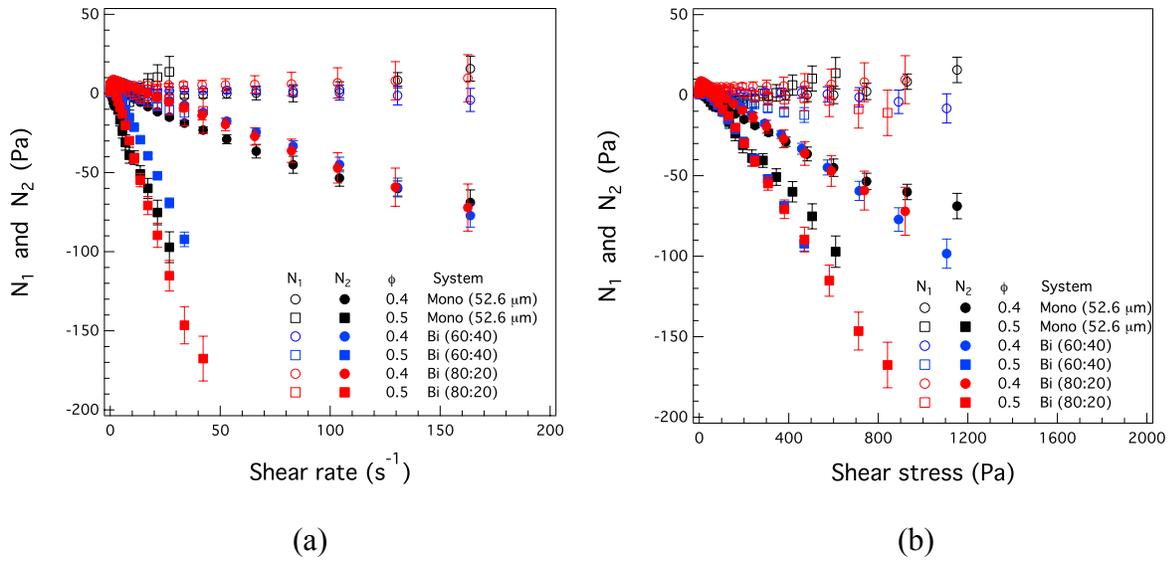

FIG. 8. Normal stress differences $N_1$ and $N_2$ of mono- and bidisperse suspensions at volume fractions of 0.4 and 0.5 as functions of (a) shear rate and (b) shear stress.

The ratio $-N_2/\tau$ for the mono- and bidisperse suspensions is plotted versus volume fraction in Fig. 9, where the shear stresses used are greater than 100 Pa, together with results of Zarraga, Hill, and Leighton [2000]; Singh and Nott [2003]; Couturier *et al.* [2011] and Dai *et al.* [2013]. The scaling $N_2/\tau = -4.4\phi^3$ suggested by Dai *et al.* [2013] is also shown. There is a reasonable agreement with Zarraga, Hill, and Leighton [2000] for the monodisperse system of 10 μm particles, but the 52.6 μm monodisperse suspensions and both bidisperse suspensions lie well below the other data. Values of $N_2/\tau$ for both bidisperse systems were comparable; for the $\phi$ = 0.4 suspension the values were close to those for the 52.6 μm monodisperse system.

$N_1/\tau$ is shown as a function of volume fraction in Fig. 10. There is a large amount of scatter, and it is not possible to draw any concrete conclusions, other than that the magnitude is small and close to zero. The bidisperse systems at $\phi = 0.5$ appear to have a sign opposite that of the monodisperse systems, but the uncertainty in both sets of calculations is too large to have any confidence in this result.

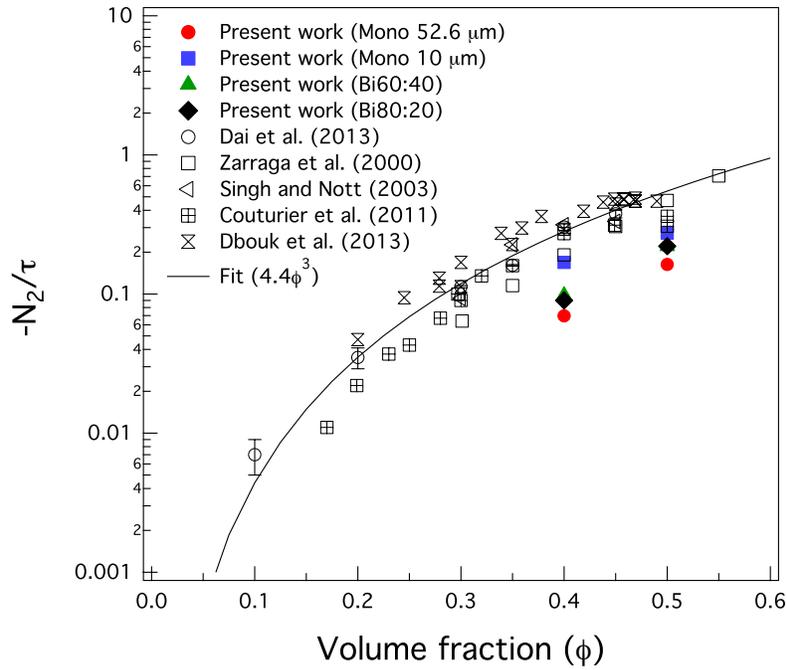

FIG. 9. Comparison of present results for $-N_2/\tau$ with experimental results from other studies.

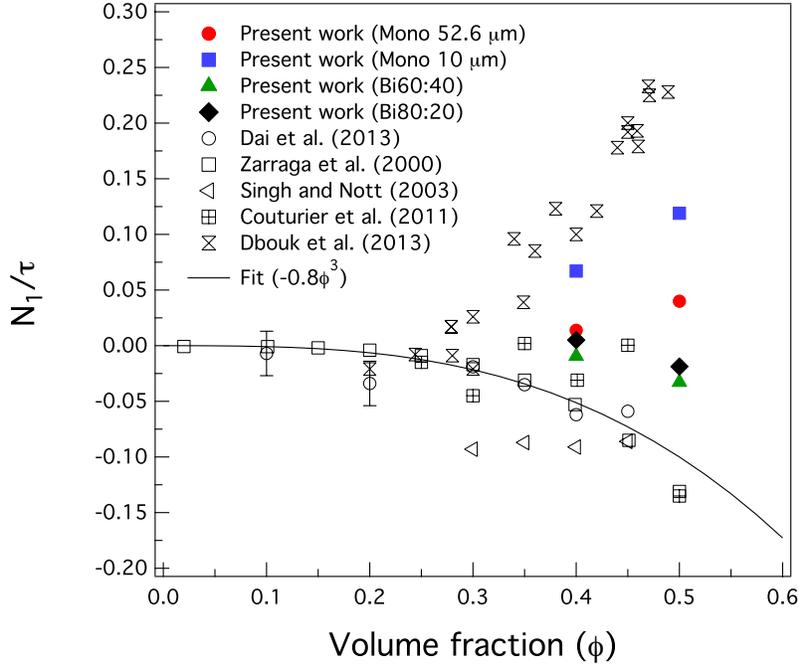

FIG. 10. Comparison of present results for $N_1/\tau$ with experimental results from other studies.

IV. CONCLUSION

This work reaffirms that $N_2$ is negative for non-Brownian suspensions of spheres and larger in magnitude than $N_1$, and, in qualitative agreement with Dbouk, Lobry, and Lemaire [2013], we find that $N_1$ is small but positive for our mono disperse suspensions. The effect of bidispersity is to shift the predicted jamming transition to higher volume fractions and hence reduce the magnitude of the viscosity at a given volume fraction. The normal stresses are insensitive to bidispersity when plotted versus the shear stress. The method of Marsh and Pearson appears to be an effective means of using total force measurements to determine $N_1$ and $N_2$ in suspensions for which the particles are too large for conventional cone-and-plate rheometry. The strong particle size dependence for the PMMA spheres in the aqueous Newtonian surfactant suspending fluid is unexpected in light of prior studies and points to an effect of fluid-particle surface chemistry that has previously been neglected.

**ACKNOWLEDGMENTS**

Chaiwut Gamonpilas is on leave from MTEC. Useful discussions with Dr. Romain Mari are gratefully acknowledged.

**Appendix: Glass suspensions**

We have performed additional viscosity measurements on glass suspensions using two particle sizes: 41.5 ± 3.6 μm and 122.6 ± 8.7 μm at concentrations from $\phi$ = 0.2 to 0.5. The Newtonian suspending fluid was a mixture of 70 wt% corn syrup and 30 wt% glycerine, similarly to that used in the work of Zarraga, Hill, and Leighton [2000], with a viscosity of 2.62 ± 0.16 Pas between shear rates of 0.5 and 500 $s^{-1}$. The relative viscosities of the glass suspensions are shown in Fig. 11 for $\phi$ = 0.2, 0.4 and 0.46. As with the PMMA suspension, Newtonian behavior was observed at low solid fraction, whereas shear thinning was apparent at higher solid loadings. Experimental reproducibility was good for both particle sizes, although substantial scatter was observed at shear rates below about 1 $s^{-1}$ for the most concentrated suspension. There is a small particle-size dependence at the higher volume fractions, but notably it is the opposite of that observed for the PMMA suspensions, in that the *larger* particles show a higher relative viscosity.

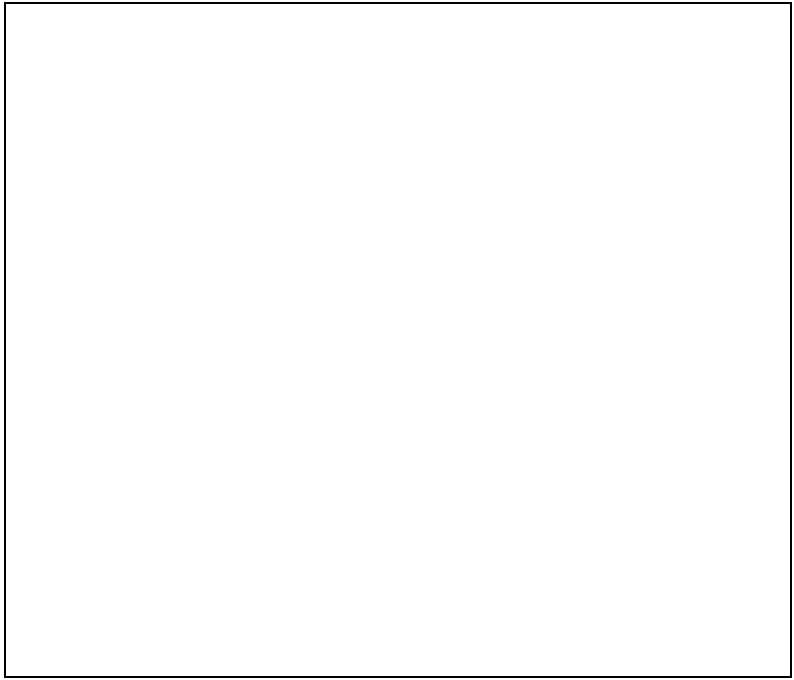

FIG. 11. Relative viscosities of glass sphere suspensions for two sizes of particles (41.5 ± 3.6 and 122.6 ± 8.7 μm).

The relative viscosity as a function of particle fraction at a shear rate of 0.1 s$^{-1}$ is shown in Fig. 12, together with other experimental studies on glass particles. The results are consistent with previous investigations by Lewis and Nielsen [1968], Shapiro and Probstein [1992], and Zarraga, Hill, and Leighton [2000], and they are bounded by the two latter studies at the higher volume fractions.

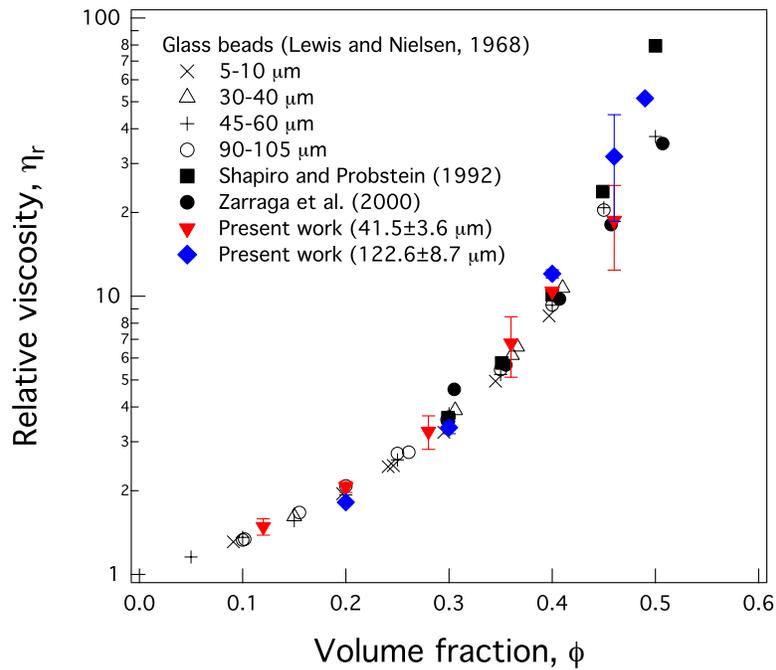

FIG. 12. Relative viscosity of suspensions of glass spheres as a function of particle fraction. Comparison of the present measurements at 0.1 s$^{-1}$ with the results of Lewis and Nielsen [1968], Shapiro and Probstein [1992], and Zarraga, Hill, and Leighton [2000].